\documentclass[12pt]{article}

\input epsf
\usepackage{amssymb,amsmath,subfigure,float}
\usepackage[matrix,arrow,curve]{xy}
\usepackage{cite}
\usepackage{multirow}
\usepackage{graphicx,color}
\usepackage[debug,pageanchor=false]{hyperref}
\definecolor{link}{rgb}{.8,.15,.1}
\hypersetup{colorlinks=true,linkcolor=link,citecolor=link,urlcolor=link,linktocpage}

\makeatletter
\@addtoreset{equation}{section}
\makeatother

\begin{document}

\begin{titlepage}

\begin{center}

\vskip .3in \noindent

{\Large \bf{Dimensional reduction of BPS attractors in AdS gauged supergravities \\\vspace{.2cm}}}

\bigskip

	Kiril Hristov\\

       \bigskip
		 Dipartimento di Fisica, Universit\`a di Milano--Bicocca, I-20126 Milano, Italy\\
       and\\
       INFN, sezione di Milano--Bicocca,
       I-20126 Milano, Italy

       \vskip .5in
           
       {\bf Abstract }
        \vskip .1in        
        \end{center}         
       
	We relate across dimensions BPS attractors of black strings and black holes of various topology in gauged supergravities with nontrivial scalar potential. The attractors are of the form AdS$_{2, 3} \times \Sigma^{2, 3}$ in 4, 5, and 6 dimensions, and can be generalized to some higher dimensional analogs. Even though the attractor geometries admit standard Kaluza-Klein and Scherk-Schwarz reductions, their asymptotic AdS spaces in general do not. The resulting lower dimensional objects are black holes with runaway asymptotics in supergravity theories with no maximally symmetric vacua. Such classes of solutions are already known to exist in literature, and results here suggest an interpretation in terms of their higher-dimensional origin that often has a full string theory embedding. In a particular relevant example, the relation between 5d Benini-Bobev black strings \cite{Benini:2013cda} and a class of 4d Cacciatori-Klemm black holes \cite{Cacciatori:2009iz} is worked out in full detail, providing a type IIB and dual field theory description of the latter solutions. As a consistency check, the Cardy formula for the field theory is shown to match the Bekenstein-Hawking entropy for horizon topology of any genus.
       

\noindent

\vfill
\eject

\end{titlepage}

\section{Introduction} 
\label{sec:intro}
The topic of dimensional reduction between supergravity theories and their vacua has been a very fruitful area of research for the development of string theory \cite{deWit:1992wf,Kugo:2000hn,Andrianopoli:2004im,Andrianopoli:2005jv,Gaiotto:2005gf,Gaiotto:2005xt,Behrndt:2005he,Looyestijn:2010pb,Banerjee:2011ts}. The relation between supersymmetric black (st)rings and black holes (even if not made fully explicit at the time) in ungauged supergravities in 6d/5d and 5d/4d was crucial for the microscopic understanding of black hole entropy \cite{Strominger:1996sh,Maldacena:1997de} and has therefore given us a tool to look into the quantum regime of black hole physics. Following this train of thought, here and in a companion paper \cite{HR} we explore similar relations between BPS black objects in 4, 5, and 6 dimensions, this time in gauged supergravity. In particular here we look at the dimensional reduction for gauged theories with a nontrivial scalar potential, starting from asymptotically AdS$_{5,6}$ black hole solutions and using standard Kaluza-Klein and Scherk-Schwarz reduction ans\"{a}tze.   

The question of direct dimensional reduction of full black hole spacetimes from 6 to 5 or from 5 to 4 dimensions is slightly more subtle in the presence of scalar potential than in the absense of one. It is not possible geometrically to construct a relation AdS$_6 \rightarrow$AdS$_5 \rightarrow$AdS$_4$ by declaring that one spatial dimension becomes compact and small, as in the case Mink$_6 \rightarrow$Mink$_5 \rightarrow$Mink$_4$. However, even in the presence of scalar potential one can still relate the near-horizon geometry AdS$_3 \times$S$^3$ (or a more general 3-surface $\Sigma^3$) of the 6 dimensional black string to the lower-dimensional near-horizon geometries AdS$_2 \times$S$^3$ and AdS$_3 \times$S$^2$ of the 5d black holes and strings and eventually to the 4d black hole attractor AdS$_2 \times$S$^2$ (as already done in \cite{LozanoTellechea:2002pn} in ungauged supergravity). The near-horizon geometries of the lower-dimensional solutions cannot be anymore connected by an RG flow to asymptotic AdS flows as the original ones, but rather to curved domain walls \cite{Cacciatori:2009iz}. 

This relation between attractors is rather general and can be applied to a variety of already known or yet to be found solutions in gauged supergravities of various dimensions. It can be used as an additional tool to classify supergravity solutions in theories with runaway behavior that might on first sight seem ill defined. When lifted to one extra dimension these solutions become perfectly good and should therefore be taken seriously, which is the main message of this paper.

The explicit example worked out in detail here relates two well-known classes of black objects. If we take a black string in 5d, i.e.\ a spacetime interpolating between AdS$_3 \times \Sigma^2$ and AdS$_5$, the 4d result after dimensional reduction along the circle in AdS$_3$ looks like a solution interpolating between AdS$_2 \times \Sigma^2$ and a non-maximally symmetric runaway vacuum. Such classes of solutions are already known \cite{Cacciatori:2009iz} and it is interesting to observe that such a dimensional reduction can preserve the full amount of supersymmetry. It turns out that the BPS black strings summarized in \cite{Benini:2013cda} upon dimensional reduction give a class of BPS black hole solutions with curved domain wall asymptotics found in \cite{Cacciatori:2009iz} for the 4d $N=2$ supergravity prepotential
$$F = \frac{X^1 X^2 X^3}{X^0}\ .$$ This suggests a more transparent string theory interpretation of the 4d solutions as asymptotic AdS$_5 \times$S$^5$ states in type IIB string thery with well-known field theory dual. One can even find much about the superconformal field theory living on the horizon of these solutions from RG-flow techniques developed in \cite{Benini:2013cda}. In this way the dimensional reduction allows us to transform a somewhat obscure 4d object into a full string theory background with a well-understood microscopic description. We stress that via the dimensional reduction the class of Cacciatori-Klemm (CK) solutions with the above prepotential now have the interpretation of wrapped D3 branes. These are different from the probably better known class of CK solutions embeddable in 11d supergravity that have the interpretation of wrapped M2 branes, found in an $N=2$ supergravity theory with prepotential\footnote{Here we make a distinction between two supergravity theories with prepotentials that are dual to each other in the ungauged case. However in both theories we allow only for electric gauging of the R-symmetry, therefore we break the duality and truly find two physically distinct models.}
$$F = -2 i \sqrt{X^0 X^1 X^2 X^3}\ .$$

The outline of the paper is as follows. In section \ref{sec:reduction} we discuss in more detail the general features of the reduction on a circle of theories with bare cosmological constant or scalar potential. We then go to the specific case of reduction between 5d black string and 4d black hole attractors in section \ref{sec:5to4}. This is done in full detail, discussing various aspects such as symmetry algebras and dual field theories. The explicit reduction further reveals a possible generalization of the known black string near-horizon geometries. We then move to the relation between 6d and 5d attractors in section \ref{sec:6to5}, which is sketched more briefly with an emphasis on the new features such as the choice of compactification directions. After this point further generalizations to higher dimensions should be clear to the reader, examples of which can be reductions on end points of holographic RG flows a la Maldacena-Nunez \cite{Maldacena:2000mw} that can be found in \cite{Gauntlett:2003di,Benini:2013cda} and references therein.

\section{Dimensional reduction of gauged theories}
\label{sec:reduction}
It is instructive to see in a simple example why the dimensional reduction on a circle of a theory of gauged supergravity leads to a lower dimensional supergravity theory with no maximally symmetric vacua\footnote{Note that there exist some very special examples of the opposite, see \cite{Klebanov:2004ya} for a 6d theory admitting an AdS$_5 \times$S$^1$ vacuum.}. In particular, the lower dimensional theory cannot have an AdS$_d$ vacuum if the original one had an AdS$_{d+1}$ vacuum (for $d > 2$). This is clear from geometric point of view since AdS$_{d+1}$ is not an S$^1$ fibre over AdS$_d$, except when $d=2$. The special case of reduction from AdS$_3$ to AdS$_2$ (and the analogous one from S$^3$ to S$^2$) is exactly the reason why the dimensional reduction between near-horizon geometries of black objects of various dimensions works in first place.

Let us consider the simplest gravitational theory in $(d+1)$ dimensions with a cosmological constant, such that it has an AdS$_{d+1}$ vacuum solution:
\begin{equation}\label{firstlagr}
 S_{d+1} = \int {\rm d} x^{d+1}\ \sqrt{g_{d+1}} \left( R_{d+1} + \Lambda\right)\ .
\end{equation}
We can reduce to $d-$dimensions with the most-general Kaluza-Klein (KK) ansatz,
\begin{equation}\label{KKmetric}
  {\rm d} s^2_{d+1} = e^{-2 \alpha \phi} {\rm d} s^2_d + e^{2 \beta \phi} ({\rm d}x_{d+1} + A_m {\rm d} x^m )^2\ ,
\end{equation}
where $\phi$ and $A_m$ are the KK scalar and vector, respectively. Under the above decomposition we have that $\sqrt{g_{d+1}} = e^{(\beta-d \alpha) \phi} \sqrt{g_d}$. Since we would like to remain for simplicity in the Einstein frame in the lower dimensional theory\footnote{There is no loss of generality in this choice, but the main argument here is easier to see this way.} we require that $\beta = (d-2) \alpha$. We therefore find that the lower dimensional action contains the following terms:
\begin{equation}\label{secondlagr}
 S_{d} =  \int {\rm d} x^{d}\ \sqrt{g_{d}} \left( R_{d} + \Lambda e^{-2 \alpha \phi} + ...\ \right)\ ,
\end{equation}
where the extra terms `...' are the kinetic terms for the KK scalar and vector, depending on the particular value of the parameter $\alpha$. One can see that the lower dimensional theory has a scalar potential that originates from the cosmological constant in \eqref{firstlagr},
\begin{equation}
 \label{scpotential}
 V = \Lambda e^{-2 \alpha \phi}\ ,
\end{equation}
which has no critical points. In fact the potential has a runaway behavior with a global extremum at $\phi \rightarrow \infty$, which is the decompactification limit. The gravitational ground state of the lower dimensional theory is in a sense trying to come back to its higher dimensional AdS$_{d+1}$ origin. If one is only given the lower dimensional lagrangian \eqref{secondlagr} the conclusion would be that there is no ground state, and all solutions of the theory asymptote to a sort of curved domain wall (see more comments in section 3.3 of \cite{Cacciatori:2009iz}). This feature is completely general and one can easily substitute the plain cosmological constant $\Lambda$ in \eqref{firstlagr} with a more complicated scalar potential that generically appears in supergravity theories. The appearance of the extra KK scalar from the metric reduction always means that the lower-dimensional scalar potential has no local extrema in the extra scalar direction, as we see in the folowing sections.   

\section{5d to 4d: black string $\rightarrow$ black hole}
\label{sec:5to4}
In order to show our main point and connect two classes of known solutions in 4 and 5 dimensional supergravities\footnote{The relation between 4d and 5d flow equations for extremal black branes in gauged supergravity was already noted in \cite{BarischDick:2012gj}. Here we take a different approach and relate already known BPS solutions, rather than the underlying equations.}, we start from the Benini-Bobev (BB) black string solutions \cite{Benini:2013cda}. They are quarter-BPS numerical solutions interpolating between AdS$_3 \times \Sigma^2$ and asymptotically locally AdS$_5$ foliated in $\mathbb{R} \times S^1\times\Sigma^2$ coordinates ($\Sigma^2$ can be any compact Riemann surface). We already discussed above that one cannot expect a proper reduction to 4 dimensions of the full flow, therefore we concentrate in particular on the near-horizon reduction from AdS$_3 \times \Sigma^2$ down to AdS$_2 \times \Sigma^2$. Once this is done we can show that the resulting 4 dimensional geometry is still supersymmetric\footnote{Note that there is another possible supersymmetric reduction from AdS$_3 \times \Sigma^2$ down to AdS$_3$ in 3d gauged supergravity \cite{Karndumri:2013iqa}. It provides further understanding of the gravitational counterpart of c-extremization.} and has an asymptotic flow to a runaway vacuum, exactly as expected. This suggests that from a full 10-dimensional perspective one should rather think of the two solutions as being the same, the difference arising only when the extra dimensions are forgotten.

\subsection{Benini-Bobev black strings}
\label{subsec:BB}
We consider $N=2$ $D=5$ Fayet-Iliopoulos (FI) gauged supergravity with two vector multiplets (the so-called STU model), which is a truncation of the maximal $N=8$ supergravity arising from the compactification of type IIB supergravity on S$^5$. The bosonic fields are the metric $g_{\mu \nu}$, 2 real scalars $\phi^{1,2}$, and three abelian gauge fields $A^{1,2,3}_{\mu}$. The bosonic lagrangian in standard conventions \cite{Klemm:2000nj} is given by
\begin{eqnarray}
e^{-1}{\cal {L}} &=&\frac{1}{2}R+g^{2}V-{\frac{1}{4}}G_{IJ}F_{\mu \nu
}{}^{I}F^{\mu \nu J}-\frac{1}{2}{\cal G}_{ij}\partial _{\mu }\phi
^{i}\partial ^{\mu }\phi ^{j}  \nonumber \\
&&+{\frac{e^{-1}}{48}}\epsilon ^{\mu \nu \rho \sigma \lambda }C_{IJK}F_{\mu
\nu }^{I}F_{\rho \sigma }^{J}A_{\lambda }^{K},  \label{action}
\end{eqnarray}
with a gauge coupling constant $g$ and a scalar potential given by 
\begin{equation}\label{scal_pot5}
V(X)=V_{I}V_{J}\left( 6X^{I}X^{J}-{\frac{9}{2}}{\cal G}^{ij}\partial
_{i}X^{I}\partial _{j}X^{J}\right) , 
\end{equation}
where $X^{I}$ represent the real scalar fields and satisfy the condition ${\cal V}={\frac{1}{6}}C_{IJK}X^{I}X^{J}X^{K}=1$. In the gauged STU model we further have $V_1 = V_2 = V_3 = \frac{1}{3}$ and $C_{1 2 3} = 1$ and its permutations as only nonvanishing components.
The physical quantities in \eqref{action} can be expressed in terms of the homogeneous cubic
polynomial ${\cal V}$.
We also have the relations 
\begin{eqnarray}
G_{IJ} &=&-{\frac{1}{2}}\partial _{I}\partial _{J}\log {\cal V}\Big|_{{\cal V%
}=1},  \nonumber \\
{\cal G}_{ij} &=&\partial _{i}X^{I}\partial _{j}X^{J}G_{IJ}\Big|_{{\cal V}%
=1},
\end{eqnarray}
where $\partial _{i}$ and $\partial _{I}$ refer, respectively, to a partial
derivative with respect to the scalar field $\phi^{i}$ and $%
X^{I}=X^{I}(\phi ^{i})$.

Note that the only difference between gauged and ungauged supergravity at the level of the presented bosonic lagrangian is the appearance of the scalar potential \eqref{scal_pot5}.

The near-horizon solutions in \cite{Benini:2013cda}, that summarize the earlier work of \cite{Cucu:2003bm,Cucu:2003yk,Naka:2002jz,Gauntlett:2006qw,Almuhairi:2011ws}, are given by
\begin{equation}
 {\rm d} s^2 = R^2_{AdS_3}\ {\rm d} s^2_{AdS_3} + R^2_{\Sigma}\ {\rm d} \sigma^2_{\Sigma}\ ,
\end{equation}
\begin{equation}
 F^I = - a_I\ Vol_{\Sigma}\ ,
\end{equation}
\begin{equation}
 X^1 = e^{-\frac{\phi^1}{\sqrt{6}}-\frac{\phi^2}{\sqrt{2}}}, \quad X^2 = e^{-\frac{\phi^1}{\sqrt{6}}+\frac{\phi^2}{\sqrt{2}}}, \quad X^3 = e^{ \frac{2 \phi^1}{\sqrt{6}}}\ .
\end{equation}
The solution is fully determined by the magnetic charges $a_I$ that satisfy the constraint 
\begin{equation}\label{dirac}
 a_1+a_2+a_3 = - \kappa, 
\end{equation}
with $\kappa = +1, -1,$ or $0$ depending on the curvature of the Riemann surface $\Sigma$ being positive, negative or zero. We have
\begin{equation}
  e^{\sqrt{6} \phi^1} = \frac{a_3^2 (a_1+a_2-a_3)^2}{a_1 a_2 (-a_1+a_2+a_3) (a_1-a_2+a_3)}, \quad e^{\sqrt{2} \phi^2} = \frac{a_2 (a_1-a_2+a_3)}{a_1 (-a_1+a_2+a_3)}\ ,
\end{equation}
\begin{equation}
R_{AdS_3}^3 = \frac{8 a_1 a_2 a_3 \Pi}{\Theta^3}, \quad R_{\Sigma}^6 = \frac{a_1^2 a_2^2 a_3^2}{\Pi}
\end{equation}
with
$$\Pi = (a_1+a_2-a_3) (a_1-a_2+a_3) (-a_1+a_2+a_3), \quad \Theta = 2 (a_1 a_2+a_1 a_3+a_2 a_3) - (a_1^2+a_2^2+a_3^2)\ .$$
We therefore find 
\begin{equation}
 X^1 = \frac{a_1 (-a_1+a_2+a_3)}{(a_1 a_2 a_3 \Pi)^{1/3}}, \quad X^2 = \frac{a_2 (a_1-a_2+a_3)}{(a_1 a_2 a_3 \Pi)^{1/3}}, \quad X^3 = \frac{a_3 (a_1+a_2-a_3)}{(a_1 a_2 a_3 \Pi)^{1/3}}\ .
\end{equation}
We can choose to write the metric of AdS$_3$ locally in the extremal BTZ form
\begin{equation}\label{btz}
 {\rm d} s^2_{AdS_3} = \frac{1}{4} \left( - r^2\ {\rm d} t^2 + \frac{{\rm d} r^2}{r^2} \right) + \rho_+ \left( {\rm d} y + \left( -\frac{1}{4} +\frac{r}{2 \rho_+} \right) {\rm d} t \right)^2\ ,
\end{equation}
with an arbitrary constant $\rho_+$ that corresponds to the horizon radius of the extremal BTZ black hole such that upon reduction we have a more general solution with an arbitrary Kaluza-Klein electric charge, as done standardly in literature. Strictly speaking, the global properties of this metric are different than those of pure AdS$_3$. However locally the metrics are the same, therefore the equations of motion and BPS variations are not sensitive under this change in the metric.

\subsection{Caciatori-Klemm black holes}
\label{subsec:CK}
Now we consider the analogous STU model, but in $N=2$ $D=4$ FI gauged supergravity with three vector multiplets. The bosonic fields are the metric $g_{\mu \nu}$, 3 complex scalars $z^{1,2,3}$, and four abelian gauge fields $A^{0,1,2,3}_{\mu}$. In standard $N=2$ conventions (see \cite{Andrianopoli:1996cm}), the lagrangian and susy variations are given in terms of symplectically covariant vectors such as to make the duality group manifest. The lagrangian is further uniquely defined by specifying the FI parameters $\xi_1=\xi_2=\xi_3 = 1$ and arbitrary $\xi_0$, and prepotential 
\begin{equation}\label{prepot}
 F = \frac{X^1 X^2 X^3}{X^0}\ ,
\end{equation}
where the $X^{\Lambda} (z^i)$ are the holomorphic sections of special geometry. With this choice of prepotential and FI parameters, the scalar potential is independent of $\xi_0$. For vanishing real part of the complex scalars, one finds the potential
\begin{equation}
 V (z) \sim \frac{1}{{\rm Im} z^1} + \frac{1}{{\rm Im} z^2} + \frac{1}{{\rm Im} z^3}\ , 
\end{equation}
which has no extrema and a typical runaway behavior exactly as explained in the previous section.

The Cacciatori-Klemm solutions and their generalizations \cite{Cacciatori:2009iz,Dall'Agata:2010gj,Hristov:2010ri,Halmagyi:2013qoa,Halmagyi:2013uza,Katmadas:2014faa,Halmagyi:2014qza} in this theory can be found from the following equations\footnote{Here we only look for axion-free solutions and use the conventions in \cite{Katmadas:2014faa}. Their generalizations with axions relate also to more general 5d solutions of black strings with additional electric charges, see section \ref{subsec:extension}.} 
\begin{eqnarray}
  \nonumber \beta^1 (-\beta^1+\beta^2+\beta^3) = p^1\ ,\\
  \nonumber \beta^2 (\beta^1-\beta^2+\beta^3) = p^2\ ,\\
  \beta^3 (\beta^1+\beta^2-\beta^3) = p^3\ ,\\
  \nonumber \beta_0 (\beta^1 + \beta^2 + \beta^3) = q_0\ ,
\end{eqnarray}
where the full solution for the metric and scalars is given by constants $\beta_0, \beta^{1,2,3}$ as in \cite{Katmadas:2014faa} and $q_0$ and $p^{1,2,3}$ are non-vanishing electric and magnetic charges carried by the gauge fields. We can already jump ahead of the dimensional reduction and adopt the notation where $p^1 \equiv a_1, p^2 \equiv a_2, p^3 \equiv a_3$. Supersymmetry again imposes \eqref{dirac}. The solution of the above equations is then
\begin{equation}
 \beta_0 = q_0 \frac{\sqrt{\Pi}}{\Theta}, \quad \beta^1 = \frac{a_1 (-a_1+a_2+a_3)}{\sqrt{\Pi}}, \quad ...
\end{equation}
with cyclic permutations of the indices for $\beta^{2,3}$. Two other useful identities are
\begin{equation}
 \beta^1+\beta^2+\beta^3 = \frac{\Theta}{\sqrt{\Pi}}, \qquad \beta^1 \beta^2 \beta^3 = \frac{a_1 a_2 a_3}{\sqrt{\Pi}}\ .
\end{equation}
We can now in principle write down the full solution that flows between a near-horizon geometry and a runaway vacuum at infinity, called a curved domain wall in \cite{Cacciatori:2009iz}. For the purposes of the explicit dimensional reduction we only concentrate on the horizon, since the asymptotic geometry is already of the expected type that suggests its higher dimensional origin as explained in section \ref{sec:reduction}. One finds the standard AdS$_2 \times \Sigma^2$ metric with different radii and scalars given by
 \begin{equation}
\tilde{R}^2_{AdS_2} = 2 \frac{\sqrt{a_1 a_2 a_3 q_0}\ \Pi}{\Theta^{5/2}}, \quad \tilde{R}^2_{\Sigma} = 2 \frac{\sqrt{a_1 a_2 a_3 q_0}}{\sqrt{\Theta}}\ ,
\end{equation}
\begin{equation}
 {\rm Im} z^1 = \frac{\sqrt{q_0}}{\sqrt{a_1 a_2 a_3 \Theta}} a_1 (-a_1+a_2+a_3), \quad ...
\end{equation}
with cyclic permutation for $z^{2,3}$.

Furthermore, observe that non-vanishing $q_0$ means that there is a nonvanishing electric component of the gauge field $A^0 = \tilde{q}^0 r\ {\rm d} t$
 with
\begin{equation}
 \tilde{q}^0 = - \frac{\tilde{R}_{AdS_2}}{2 \tilde{R}_{\Sigma}^2} \mathcal{I}^{-1, 0 0} q_0 = \sqrt{\frac{a_1 a_2 a_3}{4 q_0 \Theta}}\ ,
\end{equation}
where $\mathcal{I}^{-1}$ is the inverse of the imaginary part of the period matrix of special geometry (see again \cite{Andrianopoli:1996cm} for more details),
$$\mathcal{I}^{-1, 00} = -\frac{1}{{\rm Im} (z^1 z^2 z^3)}\ .$$

\subsection{Dimensional reduction}
\label{subsec:dimred}
We now follow the standard rules of dimensional reduction from 5d to 4d supergravity, derived in \cite{Gaiotto:2005gf,Gaiotto:2005xt,Behrndt:2005he,Banerjee:2011ts}. Note that we already used the prepotential \eqref{prepot} in 4d that one finds from reducing the 5d theory via the formula
$$F = \frac{1}{6} \frac{C_{I J K} X^I X^J X^K}{X^0}\ ,$$
together with the FI terms $\xi_1=\xi_2=\xi_3$. We did not specify $\xi_0$ as it can be left arbitrary also during the reduction - it is zero for a Kaluza-Klein reduction, or arbitrary non-zero constant for a more general Scherk-Schwarz reduction (see more in \cite{HR} or \cite{Andrianopoli:2004im,Andrianopoli:2005jv,Looyestijn:2010pb}). It is a further consistency check to see that the reduction goes through independent of the value of $\xi_0$. 

The rules for reducing the bosonic fields from 5 to 4 dimensions are the following:
\begin{equation}\label{rule1}
 {\rm d} s^2_5 = e^{2 \phi} {\rm d} s^2_4 + e^{-4 \phi} ({\rm d} \gamma - A^0_4)^2\ ,
\end{equation}
\begin{equation}\label{rule2}
 A^I_5 = A^I_4 + {\rm Re} z^I ({\rm d} \gamma - A^0_4)\ ,
\end{equation}
\begin{equation}\label{rule3}
 X^I_5 = 2 e^{2 \phi}\ {\rm Im} z^I\ ,
\end{equation}
where the 4d fields are already in the standard 4d $N=2$ conventions. Reducing the 5d solution along $\gamma$ and identifying $e^{- 2 \phi} = \rho_+ R_{AdS_3}$, we end up with a 4d solution of the form
\begin{equation}
 {\rm d} s^2 = \frac{\rho_+ R^3_{AdS_3}}{4} {\rm d} s^2_{AdS_2} + R^2_{\Sigma}\ \rho_+ R_{AdS_3} {\rm d} \sigma^2_{\Sigma}\ ,
\end{equation}
\begin{equation}
 F^0 = \frac{1}{2 \rho_+}\ Vol_{AdS_2}, \quad F^I = - a_I\ Vol_{\Sigma}\ ,
\end{equation}
\begin{equation}
 {\rm Im}z^I = \frac{\rho_+ R_{AdS_3}}{2} X^I\ .
\end{equation}
Therefore we expect an exact match between the following quantities upon the 5d-4d reduction:
 \begin{equation}
\tilde{R}^2_{AdS_2} = \frac{\rho_+ R^3_{AdS_3}}{4}, \quad \tilde{R}^2_{\Sigma} = R^2_{\Sigma}\ \rho_+ R_{AdS_3}\ ,
\end{equation}
together with the match of the scalars $z^I$ and field strengths. We have one arbitrary constant $\rho_+$ that needs to reproduce correctly all these quantities and it turns out everything agrees exactly upon the identification
\begin{equation}
 \rho_+ = \sqrt{\frac{q_0 \Theta}{a_1 a_2 a_3}}\ .
\end{equation}
This concludes the proof that the near-horizon geometries of the BB solutions reduce to the near-horizon geometries of the CK solutions for the prepotential \eqref{prepot}. More generally, this suggests that the full 4d flow of the CK solutions between horizon and a curved domain wall is to be seen from a higher dimensional perspective as a black string in AdS$_5$. In other words, the CK solutions of the 4d gauged supergravity defined by the prepotential \eqref{prepot} are to be identified with the BB solutions and thus embedded in an asymptotically AdS$_5 \times$S$^5$ background of type IIB string theory.

\subsection{Matching asymptotics}
\label{subsec:asymp}
To provide further evidence for the statement that the full CK solutions lift to the BB solutions in 5d, we can show how the asymptotic geometries match exactly. It would of course be desirable to show the full lift from the rules \eqref{rule1}-\eqref{rule3}, but this is practically not possible since the BB solutions are known only numerically. One can instead hope in this way to discover the analytic form of the BB black strings in 5d, but this is not immediately obvious as it involves the construction of the additional function $e^{2 \phi}$ and we leave it for future investigations. 

Concentrating only at the asymptotics, we find that the CK solutions we are looking at have the following runaway behavior for the metric and scalars at infinity (c.f.\ section 3.3 of \cite{Cacciatori:2009iz}):
\begin{equation}
 {\rm d} s^2_{4, y \rightarrow \infty} = y^{3/2} \left( -{\rm d} t^2 + {\rm d} \sigma^2_{\Sigma} \right) + y^{-3/2} {\rm d} y^2, \quad {\rm Im} z^i_{y \rightarrow \infty} \sim y^{1/2}\ .  
\end{equation}
Upon the coordinate change $r = y^{1/2}$, we find
\begin{equation}
 {\rm d} s^2_{4, r \rightarrow \infty} = r^3 \left( -{\rm d} t^2 + {\rm d} \sigma^2_{\Sigma} \right) + r^{-1} {\rm d} r^2, \quad {\rm Im} z^i_{r \rightarrow \infty} \sim r\ ,  
\end{equation}
which can easily be uplifted to AdS$_5$ with constant scalars via \eqref{rule1}-\eqref{rule3} upon identifying $e^{2 \phi}_{r \rightarrow \infty} = r^{-1}$,
\begin{equation}
 {\rm d} s^2_{5, r \rightarrow \infty} = r^2 \left( -{\rm d} t^2 + {\rm d} \sigma^2_{\Sigma} \right) + r^{-2} {\rm d} r^2 + r^2 {\rm d} \gamma^2, \quad X^I_{r \rightarrow \infty} \sim const\ .  
\end{equation}
As already announced, this is exactly the asymptotic form of the numerical BB black strings \cite{Benini:2013cda}.

\subsection{Superalgebras and Killing spinors}
\label{subsec:susy}
There is another simple reason why the reduction and BPS properties had to work out correctly. Both the BB and CK solutions are 1/4 BPS, preserving 2 supercharges, with a susy enhancement near the horizon to 4 supercharges. Furthermore, in both cases Killing spinors obey
\begin{equation}
  \epsilon = \gamma_{x y} \epsilon\ ,
\end{equation}
where the $x, y$-directions denote the internal space $\Sigma$, on which the Killing spinor is constant. The full symmetry algebra on the 5d horizon is \cite{Hristov:2013xza}
\begin{equation}
 5d: SU(1,1|1) \times SU(1,1) \times G_{\Sigma}\ ,
\end{equation}
where the symmetry group of $\Sigma$ can be $SU(2), U(1)^2$ for a sphere and torus or only a discrete group for a higher genus Riemann surface. The 4d horizon has the symmetry algebra \cite{Hristov:2011ye}
\begin{equation}
 4d: SU(1,1|1) \times G_{\Sigma}\ ,
\end{equation}
i.e.\ we have the exact same fermionic symmetries, but the extra bosonic $SU(1,1)$ has disappeared, since AdS$_3$ has an $SU(1,1)^2$ isometries versus a single $SU(1,1)$ for AdS$_2$. Here lies the main reason why the reduction preserves the same amount of supersymmetry - we reduced along an isometry not only of the bosonic solution, but also of the Killing spinor. This is not guaranteed to be the always the case, see e.g.\ \cite{Duff:1997qz,Goldstein:2008fq}, and it ensured the succes of the dimensional reduction. It is crucial to find the correct reduction ansatz, but the fact that we reduced over a full bosonic and fermionic isometry meant that there existed such an ansatz preserving all the supercharges of the original solution. 

\subsection{Dual field theory and Cardy formula}
\label{subsec:cft}
Following the above reduction on the boundary, one can find the corresponding CFT dual on the 4d horizon. The theory dual to the horizon AdS$_3$ of the BB solutions is a $(2,0)$ superconformal field theory in 2d that can be derived from the compactification of twisted $N=4$ SYM on the corresponding Riemann surface $\Sigma$. The $(2,0)$ theory naturally lives on $\mathbb{R} \times$S$^1$ and upon its reduction on the spatial circle becomes a superconformal quantum mechanics. We already know the central charge of the $(2,0)$ theory, which is the Brown-Henneaux central charge
\begin{equation}\label{centralcharge}
 c = \frac{3 R_{AdS_3}}{2 G_3}\ .
\end{equation}
This result was derived independently on the dual side via c-extremization. Upon putting this CFT on a circle with momentum, c.f.\ \cite{Kraus:2006wn},  
\begin{equation}
 L_0 - \frac{c}{24} = \rho_+^2 \frac{R_{AdS_3}}{4 G_3} = \rho_+ \frac{\eta_{\Sigma} \tilde{R}_{\Sigma}^2}{2 \pi G_4}\ ,
\end{equation}
we see that the Cardy formula
\begin{equation}
 S_{Cardy} = 2 \pi \sqrt{\frac{c}{6} \left(L_0 - \frac{c}{24} \right)} = \frac{\eta_{\Sigma}}{G_4} \sqrt{\rho_+ R_{AdS_3} R_{\Sigma}^2 \tilde{R}_{\Sigma}^2} = \frac{ \eta_{\Sigma} \tilde{R}_{\Sigma}^2}{G_4}\ ,
\end{equation}
reproduces exactly\footnote{In the above derivation we used that the area of a unit Riemann surface $\Sigma$ is $4 \eta_{\Sigma}$ and depends on its genus as in (3.6) of \cite{Benini:2013cda}.} the macroscopic Bekenstein-Hawking entropy of the CK solution that is also the entropy of the BTZ black hole,
\begin{equation}
 S_{BH} = \frac{A_{\Sigma}}{4 G_4} = \frac{\eta_{\Sigma} \tilde{R}_{\Sigma}^2}{G_4} = \frac{2 \pi \eta_{\Sigma} \rho_+ R_{AdS_3} R_{\Sigma}^2}{G_5} = \frac{\pi \rho_+ R_{AdS_3}}{2 G_3} = \frac{A_{BTZ}}{4 G_3} = S_{BTZ}\ .
\end{equation}
This is a seemingly trivial AdS/CFT check, but it is important to stress that in this case we explicitly know the dual field theory and therefore have a true microscopic description of the black hole degrees of freedom. The nontrivial part of this derivation actually consists in finding the correct value of the central charge \eqref{centralcharge} directly from the field theory as done in \cite{Benini:2013cda} via c-extremization. We stress again that we have essentially embedded a class of CK solutions inside type IIB supergravity via their relation with the BB solutions, such that now they have the interpretation of wrapped D3 branes. Thus it is not a surprise to see that the entropy scales with $N^2$ instead of $N^{3/2}$ (hidden in the scaling of the Newton constant $G_3 \sim G_4 \sim G_5 \sim N^{-2}$). It also follows that the euclideanized version of the near-horizon geometry in section \ref{subsec:CK} is the gravity dual to be compared to field theory localization calculations of twisted $N=4$ SYM on T$^2 \times \Sigma^2$ (see \cite{Closset:2013sxa} for some results in this direction).

\subsection{Extensions with extra charges}
\label{subsec:extension}
It is interesting to observe that one could write more general CK solutions within the prepotential and FI terms considered in section \ref{subsec:CK}, namely solutions with nonvanishing axions and more electric charges \cite{Halmagyi:2013uza,Katmadas:2014faa,Halmagyi:2014qza}. These have no known higher dimensional origin, but following the reduction rules in section \ref{subsec:dimred} one can retrace what nonvanishing ${\rm Re}\ z^I$ terms correspond to in the higher dimensional solution. One can add extra gauge fields $$A^I_{\gamma} = w^I\ ,$$  
which are essentially Wilson lines around the circle in AdS$_3$. Upon reduction they produce non-vanishing axions and corresponding electric charges. It is interesting to observe that such Wilson lines already give nonvanishing electric charges in 5d due to the Chern-Simons term in the supergravity lagrangian \eqref{action} and the fact that we have already switched on magnetic charges. This corresponds to a generalization of the BB solutions with electric charges and will be described more carefully in \cite{HK}.

\section{6d to 5d: black string $\rightarrow$ black hole/string}
\label{sec:6to5}
The detailed example in the previous section illustrated the general principles of dimensional reduction in gauged supergravity. Other analogous examples can be explored across different dimensions, as long as there exists a suitable spatial isometry for dimensional reduction. In gauged supergravities such suitable spacetimes with AdS$_3$, S$^3$ or other circle fibrations often arise as near-horizon geometries of various black branes. As discussed above, depending on the explicit Killing spinors of the original solution we can see if the dimensionally reduced solution also preserves supersymmetry or not.

Here we sketch more briefly another example in 6d supergravity. The near-horizon geometry of BPS black strings found in \cite{Cariglia:2004kk} is AdS$_3 \times$S$^3$. This is a solution of $N=(1,0)$ $SU(2)$ gauged supergravity in 6d with a nonvanishing $SU(2)$ field strength $F^I$. The supergravity theory includes the gravity and tensor multiplets that include the metric, a scalar $\phi$ and tensor field strength $G$, and the above mentioned $SU(2)$ gauge multiplet. The supersymmetric AdS$_3 \times$S$^3$ background is given by:
\begin{equation}
 {\rm d} s^2 = R^2_1\ {\rm d} s^2_{AdS_3} + R^2_2\ {\rm d} s^2_{S^3}\ ,
\end{equation}
\begin{equation}
G = \frac{2 (1-a^2)}{g^2 R_1 R_2^2} (Vol_{R_1} (AdS_3) + Vol_{R_2} (S^3))\ ,
\end{equation}
\begin{equation}
 F^I = \frac{(1-a^2)}{g} \epsilon^{I J K} e^{J K}, \qquad e^{\sqrt{2} \phi} = \frac{g^2 R_2^2}{2 (1-a^2)}\ .
\end{equation}
In the above formulae $g$ is the gauge coupling constant, $a \equiv R_2/R_1$, $\epsilon^{IJK}$ are the $SU(2)$ structure constants and $e^{I J}$ are two forms made from the vielbein on the three-sphere (see e.g.\ \cite{Hristov:2013xza}). The supersymmetry is preserved due to the fact that the gauge field ``cancels'' the spin connection on the internal space. The solution is quarter-BPS, i.e.\ there are four conserved supercharges with a symmetry group
$$SU(1,1|1) \times SU(1,1) \times SU(2)^2\ .$$
There is a clear similarity to the 5-dimensional case, and the fact that the Killing spinor transforms only under one $SU(1,1)$ of the AdS$_3$ isometry group and is scalar under rotations on the sphere means that now we have two ways of reducing along a circle while still preserving supersymmetry. One can reduce on a circle either inside AdS$_3$ or inside S$^3$. This leads to two different BPS solutions - AdS$_2 \times$S$^3$ with $SU(2)$ gauge fields or AdS$_3 \times$S$^2$ with $U(1)$ gauge fields. 

The resulting 5d theory coming from the reduction of the $N=(1,0)$ in 6d has a $SU(2)\times U(1)$ gauging (the extra $U(1)$ can be realized by Scherk-Scwarz reduction) and is of a runaway type, i.e.\ an $N=4$ theory with extra vectormultiplets and no extrema of the scalar potential, similar to some of the theories discussed in \cite{Gunaydin:1985cu,Romans:1985ps,Hristov:2013xza}.

\subsection{Reduction along AdS$_3$}
\label{subsec:tohole}
We can again use the BTZ form of the metric \eqref{btz} and reduce the above solution along the circle. The resulting solution has the near-horizon geometry of a black hole in 5d - AdS$_2 \times$S$^3$ with the same $SU(2)$ gauge field on the sphere. We further find two $U(1)$ gauge fields carrying an electric charge - one coming from the 6d tensor field and the other being the Kaluza-Klein vector. The fermionic symmetries remain the same, but the symmetry algebra does not contain the extra $SU(1,1)$ factor found in 6d. Such a solution cannot be connected to any AdS$_5$ vacuum, but near the horizon resembles already known attractors in 5d (see e.g.\ \cite{Naka:2002jz,Hristov:2013xza} for the case of supersymmetric AdS$_2 \times$H$^3$ with an $SU(2)$ gauge field).

\subsection{Reduction along S$^3$}
\label{subsec:tostring}
To reduce along S$^3$, one can write down the metric of the sphere in the form
\begin{equation}
 {\rm d} s^2_{S^3/\mathbb{Z}_k} = \frac{R^2_2}{4} \left( \sin^2 \theta\ {\rm d} \phi^2 + {\rm d} \theta^2 + (k {\rm d} \psi + \cos \theta\ {\rm d} \phi)^2   \right)\ ,
\end{equation}
with a slight generalization to a allow for quotients with a discrete group $\mathbb{Z}_k$. The resulting solution is less surprising - it becomes the near-horizon geometry of a black string, AdS$_3 \times$S$^2$, with magnetic charges through the sphere. The symmetry algebra contains one less $SU(2)$ factor compared to the 6d case. Again, it is found in a theory with no AdS$_5$ vacuum, but near the horizon it resembles the BB solutions in the previous section. 

\subsection{Reduction along both AdS$_3$ and S$^3$}
\label{subsec:doublered}
Finally note that one can consecutively reduce along both circles in order to get a 4d supersymmetric AdS$_2 \times$S$^2$ solution. In this case already the 5d supergravity is of runaway type and this continues to hold in 4d. The remaining supersymmetry algebra of this solution is $SU(1,1|1)\times SU(2)$ and can also be found within the near-horizon geometries of CK black holes and their generalizations.

\section*{Acknowledgments}
I would like to thank S.\ Katmadas, D.\ Klemm, E.\ \'{O} Colg\'{a}in,  A.\ Rota, A.\ Tomasiello,  A.\ Veliz-Osorio, A.\ Zaffaroni, and the JHEP referee for enlightening discussions. I am supported in part by INFN, by the MIUR-FIRB grant RBFR10QS5J ``String Theory and Fundamental Interactions'', and by the MIUR-PRIN contract 2009-KHZKRX.

\providecommand{\href}[2]{#2}

\begin{thebibliography}{10}

\bibitem{Benini:2013cda}
  F.~Benini and N.~Bobev,
  ``Exact two-dimensional superconformal R-symmetry and c-extremization,''
  Phys.\ Rev.\ Lett.\  {\bf 110} (2013) 6,  061601
  [arXiv:1211.4030 [hep-th]], 
  ``Two-dimensional SCFTs from wrapped branes and c-extremization,''
  JHEP {\bf 1306} (2013) 005
  [arXiv:1302.4451 [hep-th]].

\bibitem{Cacciatori:2009iz}
  S.~L.~Cacciatori and D.~Klemm,
  ``Supersymmetric AdS(4) black holes and attractors,''
  JHEP {\bf 1001} (2010) 085
  [arXiv:0911.4926 [hep-th]].
  
\bibitem{deWit:1992wf}
  B.~de Wit, F.~Vanderseypen and A.~Van Proeyen,
  ``Symmetry structure of special geometries,''
  Nucl.\ Phys.\ B {\bf 400} (1993) 463
  [hep-th/9210068].
  
\bibitem{Kugo:2000hn}
  T.~Kugo and K.~Ohashi,
  ``Supergravity tensor calculus in 5-D from 6-D,''
  Prog.\ Theor.\ Phys.\  {\bf 104} (2000) 835
  [hep-ph/0006231].
  
\bibitem{Andrianopoli:2004im}
  L.~Andrianopoli, S.~Ferrara and M.~A.~Lledo,
  ``Scherk-Schwarz reduction of D = 5 special and quaternionic geometry,''
  Class.\ Quant.\ Grav.\  {\bf 21} (2004) 4677
  [hep-th/0405164].
  
\bibitem{Andrianopoli:2005jv}
  L.~Andrianopoli, M.~A.~Lledo and M.~Trigiante,
  ``The Scherk-Schwarz mechanism as a flux compactification with internal torsion,''
  JHEP {\bf 0505} (2005) 051
  [hep-th/0502083].
  
\bibitem{Gaiotto:2005gf}
  D.~Gaiotto, A.~Strominger and X.~Yin,
  ``New connections between 4-D and 5-D black holes,''
  JHEP {\bf 0602} (2006) 024
  [hep-th/0503217].
  
\bibitem{Gaiotto:2005xt}
  D.~Gaiotto, A.~Strominger and X.~Yin,
  ``5D black rings and 4D black holes,''
  JHEP {\bf 0602} (2006) 023
  [hep-th/0504126].
  
\bibitem{Behrndt:2005he}
  K.~Behrndt, G.~Lopes Cardoso and S.~Mahapatra,
  ``Exploring the relation between 4-D and 5-D BPS solutions,''
  Nucl.\ Phys.\ B {\bf 732} (2006) 200
  [hep-th/0506251].
  
\bibitem{Looyestijn:2010pb}
  H.~Looyestijn, E.~Plauschinn and S.~Vandoren,
  ``New potentials from Scherk-Schwarz reductions,''
  JHEP {\bf 1012} (2010) 016
  [arXiv:1008.4286 [hep-th]].
  
\bibitem{Banerjee:2011ts}
  N.~Banerjee, B.~de Wit and S.~Katmadas,
  ``The Off-Shell 4D/5D Connection,''
  JHEP {\bf 1203} (2012) 061
  [arXiv:1112.5371 [hep-th]].
  
\bibitem{Strominger:1996sh}
  A.~Strominger and C.~Vafa,
  ``Microscopic origin of the Bekenstein-Hawking entropy,''
  Phys.\ Lett.\ B {\bf 379} (1996) 99
  [hep-th/9601029].
  
\bibitem{Maldacena:1997de}
  J.~M.~Maldacena, A.~Strominger and E.~Witten,
  ``Black hole entropy in M theory,''
  JHEP {\bf 9712} (1997) 002
  [hep-th/9711053].
  
  \bibitem{HR}
   K.~Hristov and A.~Rota,
  ``6d-5d-4d reduction of BPS attractors in flat gauged supergravities,''
  arXiv:1410.5386 [hep-th].
  
  
\bibitem{LozanoTellechea:2002pn}
  E.~Lozano-Tellechea, P.~Meessen and T.~Ortin,
  ``On d = 4, d = 5, d = 6 vacua with eight supercharges,''
  Class.\ Quant.\ Grav.\  {\bf 19} (2002) 5921
  [hep-th/0206200].
  
\bibitem{Maldacena:2000mw}
  J.~M.~Maldacena and C.~Nunez,
  ``Supergravity description of field theories on curved manifolds and a no go theorem,''
  Int.\ J.\ Mod.\ Phys.\ A {\bf 16} (2001) 822
  [hep-th/0007018].
  
\bibitem{Gauntlett:2003di}
  J.~P.~Gauntlett,
  ``Branes, calibrations and supergravity,''
  hep-th/0305074.
  
\bibitem{Klebanov:2004ya}
  I.~R.~Klebanov and J.~M.~Maldacena,
  ``Superconformal gauge theories and non-critical superstrings,''
  Int.\ J.\ Mod.\ Phys.\ A {\bf 19} (2004) 5003
  [hep-th/0409133].
  
\bibitem{BarischDick:2012gj}
  S.~Barisch-Dick, G.~Lopes Cardoso, M.~Haack and S.~Nampuri,
  ``Extremal black brane solutions in five-dimensional gauged supergravity,''
  JHEP {\bf 1302} (2013) 103
  [arXiv:1211.0832 [hep-th]].
  
\bibitem{Karndumri:2013iqa}
  P.~Karndumri and E.~O Colgain,
  ``Supergravity dual of $c$-extremization,''
  Phys.\ Rev.\ D {\bf 87} (2013) 10,  101902
  [arXiv:1302.6532 [hep-th]], 
  ``3D Supergravity from wrapped D3-branes,''
  JHEP {\bf 1310} (2013) 094
  [arXiv:1307.2086].
  
\bibitem{Klemm:2000nj}
  D.~Klemm and W.~A.~Sabra,
  ``Supersymmetry of black strings in D = 5 gauged supergravities,''
  Phys.\ Rev.\ D {\bf 62} (2000) 024003
  [hep-th/0001131].
  
\bibitem{Cucu:2003bm}
  S.~Cucu, H.~Lu and J.~F.~Vazquez-Poritz,
  ``A Supersymmetric and smooth compactification of M theory to AdS(5),''
  Phys.\ Lett.\ B {\bf 568} (2003) 261
  [hep-th/0303211].
  
\bibitem{Cucu:2003yk}
  S.~Cucu, H.~Lu and J.~F.~Vazquez-Poritz,
  ``Interpolating from AdS(D-2) x S**2 to AdS(D),''
  Nucl.\ Phys.\ B {\bf 677} (2004) 181
  [hep-th/0304022].
  
\bibitem{Naka:2002jz}
  M.~Naka,
  ``Various wrapped branes from gauged supergravities,''
  hep-th/0206141.
  
\bibitem{Gauntlett:2006qw}
  J.~P.~Gauntlett, O.~A.~P.~Mac Conamhna, T.~Mateos and D.~Waldram,
  ``New supersymmetric AdS(3) solutions,''
  Phys.\ Rev.\ D {\bf 74} (2006) 106007
  [hep-th/0608055].
  
\bibitem{Almuhairi:2011ws}
  A.~Almuhairi and J.~Polchinski,
  ``Magnetic AdS x R$^2$: Supersymmetry and stability,''
  arXiv:1108.1213 [hep-th].
  
\bibitem{Andrianopoli:1996cm}
  L.~Andrianopoli, M.~Bertolini, A.~Ceresole, R.~D'Auria, S.~Ferrara, P.~Fre and T.~Magri,
  ``N=2 supergravity and N=2 superYang-Mills theory on general scalar manifolds: Symplectic covariance, gaugings and the momentum map,''
  J.\ Geom.\ Phys.\  {\bf 23} (1997) 111
  [hep-th/9605032].
  
\bibitem{Dall'Agata:2010gj}
  G.~Dall'Agata and A.~Gnecchi,
  ``Flow equations and attractors for black holes in N = 2 U(1) gauged supergravity,''
  JHEP {\bf 1103} (2011) 037
  [arXiv:1012.3756 [hep-th]].
  
\bibitem{Hristov:2010ri}
  K.~Hristov and S.~Vandoren,
  ``Static supersymmetric black holes in AdS$_4$ with spherical symmetry,''
  JHEP {\bf 1104} (2011) 047
  [arXiv:1012.4314 [hep-th]].
  
\bibitem{Halmagyi:2013qoa}
  N.~Halmagyi,
  ``BPS Black Hole Horizons in N=2 Gauged Supergravity,''
  JHEP {\bf 1402} (2014) 051
  [arXiv:1308.1439 [hep-th]].
  
\bibitem{Halmagyi:2013uza}
  N.~Halmagyi and T.~Vanel,
  ``AdS Black Holes from Duality in Gauged Supergravity,''
  JHEP {\bf 1404} (2014) 130
  [arXiv:1312.5430 [hep-th]].
  
\bibitem{Katmadas:2014faa}
  S.~Katmadas,
  ``Static BPS black holes in U(1) gauged supergravity,''
  JHEP {\bf 1409} (2014) 027
  [arXiv:1405.4901 [hep-th]].
  
\bibitem{Halmagyi:2014qza}
  N.~Halmagyi,
  ``Static BPS Black Holes in AdS4 with General Dyonic Charges,''
  arXiv:1408.2831 [hep-th].
  
\bibitem{Hristov:2013xza}
  K.~Hristov and A.~Rota,
  ``Attractors, black objects, and holographic RG flows in 5d maximal gauged supergravities,''
  JHEP {\bf 1403} (2014) 057
  [arXiv:1312.3275 [hep-th]].
  
\bibitem{Hristov:2011ye}
  K.~Hristov, C.~Toldo and S.~Vandoren,
  ``On BPS bounds in D=4 N=2 gauged supergravity,''
  JHEP {\bf 1112} (2011) 014
  [arXiv:1110.2688 [hep-th]].
  
\bibitem{Duff:1997qz}
  M.~J.~Duff, H.~Lu and C.~N.~Pope,
  ``Supersymmetry without supersymmetry,''
  Phys.\ Lett.\ B {\bf 409} (1997) 136
  [hep-th/9704186].
  
\bibitem{Goldstein:2008fq}
  K.~Goldstein and S.~Katmadas,
  ``Almost BPS black holes,''
  JHEP {\bf 0905} (2009) 058
  [arXiv:0812.4183 [hep-th]].
  
\bibitem{Kraus:2006wn}
  P.~Kraus,
  ``Lectures on black holes and the AdS(3) / CFT(2) correspondence,''
  Lect.\ Notes Phys.\  {\bf 755} (2008) 193
  [hep-th/0609074].
  
\bibitem{Closset:2013sxa}
  C.~Closset and I.~Shamir,
  ``The $\mathcal{N}=1$ Chiral Multiplet on $T^2\times S^2$ and Supersymmetric Localization,''
  JHEP {\bf 1403} (2014) 040
  [arXiv:1311.2430 [hep-th]].
  
\bibitem{HK}
  K.~Hristov and S.~Katmadas,
  ``Wilson lines for AdS$_5$ black strings,''
  arXiv:1411.2432 [hep-th].
  
\bibitem{Cariglia:2004kk}
  M.~Cariglia and O.~A.~P.~Mac Conamhna,
  ``The General form of supersymmetric solutions of N=(1,0) U(1) and SU(2) gauged supergravities in six-dimensions,''
  Class.\ Quant.\ Grav.\  {\bf 21} (2004) 3171
  [hep-th/0402055].
  
\bibitem{Gunaydin:1985cu}
  M.~Gunaydin, L.~J.~Romans and N.~P.~Warner,
  ``Compact and Noncompact Gauged Supergravity Theories in Five-Dimensions,''
  Nucl.\ Phys.\ B {\bf 272} (1986) 598.
  
\bibitem{Romans:1985ps}
  L.~J.~Romans,
  ``Gauged $N=4$ Supergravities in Five-dimensions and Their Magnetovac Backgrounds,''
  Nucl.\ Phys.\ B {\bf 267} (1986) 433.
  
\end{thebibliography}
\end{document}